\documentclass[aps,twocolumn]{revtex4}
\usepackage{epsfig}
\newcommand{\be}{\begin{equation}}
\newcommand{\ee}{\end{equation}}
\newcommand{\bea}{\begin{eqnarray}}
\newcommand{\eea}{\end{eqnarray}}
\newcommand{\ba}{\begin{array}}
\newcommand{\ea}{\end{array}}
\newcommand{\x}{{\bf x}}

\begin{document}

\title{Tricritical directed percolation in 2+1 dimensions}
\author{Peter Grassberger}
\affiliation{John-von-Neumann Institute for Computing, Forschungszentrum
J\"ulich, D-52425 J\"ulich, Germany}

\date{\today}
\begin{abstract}
We present detailed simulations of a generalization of the Domany-Kinzel 
model to 2+1 dimensions. It has two control parameters $p$ and $q$ which 
describe the probabilities $P_k$ of a site to be wetted, if exactly $k$
of its ``upstream" neighbours are already wetted. If $P_k$ depends only 
weakly on $k$, the active/adsorbed phase transition is in the directed
percolation (DP) universality class. If, however, $P_k$ increases fast 
with $k$ so that the formation of inactive holes surrounded by active sites 
is suppressed, the transition is first order. These two transition lines
meet at a tricritical point. This point should be in the same universality
class as a tricritical transition in the contact process studied recently
by L\"ubeck. Critical exponents for it have been calculated previously by
means of the field theoretic epsilon-expansion ($\epsilon = 3-d$, with
$d=2$ in the present case). Rather poor agreement is found with either.
\end{abstract}

\maketitle

\section{Introduction}

The critical behaviour of directed percolation (DP) has been studied since more 
than 30 years, if we also count the work on `reggeon field theory' \cite{Moshe} 
which was only later recognized as a field theory for a stochastic process which 
is in the DP universality class \cite{Sunder-Grass}. In spite of this long 
history, several of its aspects have barely received any attention yet. In 
particular this is true for its tricritical version.

DP is e.g. realized by a reaction-diffusion system 
\bea
   A \to 2A \;\; && ({\rm rate}\;\; k_1),   \nonumber \\
   A \to 0  \;\; && ({\rm rate}\;\; k_2),   \nonumber \\
  2A \to A  \;\; && ({\rm rate}\;\; k_3).                   \label{r1}
\eea
Its mean field theory is just the rate equation for the number $n(t)$ of 
$A$ particles, ${\dot n} = a n - b n^2$ with $a=k_1-k_2$ and $b= k_3$. 
The transition is in this 
description at $a=0$. The first mentioning of tricritical DP (TDP) 
seems to have been made in this context by \cite{Grass82}, where it was 
pointed out that the system composed of the reactions 
\bea
  2A \to 3A \;\; ({\rm rate}\;\; k_1'),    \nonumber \\
  2A \to 0  \;\; ({\rm rate}\;\; k_2'),    \nonumber \\
  3A \to 2A \;\; ({\rm rate}\;\; k_3')                   \label{r2}
\eea
has the rate equation 
\be
   {\dot n} = b n^2 - c n^3                         \label{mft}
\ee
with $b = k_1'-2k_2', c = k_3'$. This is just the mean field equation for 
a system with tricritical point at 
\be
   b=0,\;\; c>0.                                        \label{mft2}
\ee
But, as observed by Janssen \cite{Janssen87} and Ohtsuki \& Keyes \cite{Ohtsuki}, 
Eqs.(\ref{mft}) and (\ref{mft2}) can be realized also 
differently, e.g. by combining both reaction-diffusion systems and choosing
$k_1-k_2=0, k_1'-2k_2'-k_3=0, k_3'>0$.

Both versions differ however beyond the mean field level, if fluctuations 
are taken into account. While the combination of Eqs.(\ref{r1}) and (\ref{r2})
should be described by a Langevin equation ${\dot n}(\x,t) = an(\x,t)+bn^2(\x,t)-
cn^3(\x,t) + \xi(\x,t)$ with the ``semi-multiplicative" noise, $\langle \xi(\x,t)
\xi(\x',t')\rangle \propto n(\x,t)\delta(\x-\x')\delta(t-t')$, well known from 
reggeon field theory, the noise appropriate for Eq.(\ref{r2}) alone should be 
particle number conserving when $n\to 0$, i.e. \cite{JMS}
$\langle \xi(\x,t)\xi(\x',t')\rangle \propto n^2(\x,t)\delta(\x-\x')\delta(t-t') 
+ {\rm const}\; n(\x,t)\nabla^2 \delta(\x-\x')\delta(t-t')$.

The reaction scheme Eq.(\ref{r2}) has become infamous during the last years as ``pair 
contact process with diffusion" (PCPD) \cite{Hinrich-Henkel}, and has received 
very much attention. It is still basically unsolved \cite{JWDT}. In contrast, 
there was no activity on the 
``true" tricritical DP with semi-multiplicative noise, after Ohtsuki \& Keyes
had worked out the lowest order results of the $\epsilon-$expansion. This has 
changed only very recently with field theory work by Janssen \cite{Janssen04} 
and with extensive and careful simulations by L\"ubeck \cite{Lubeck}. 

The present work was mainly started because only the stationary behaviour was
studied in \cite{Lubeck}, and we wanted to obtain also dynamical (tri-)critical
exponents. It seems that the dynamical behaviour of TDP was never studied 
numerically before. Indeed, for ordinary DP it is also easier to obtain 
static critical exponents from dynamical simulations than from stationary
behaviour. Although this might be different for TDP, it seems worth to check 
it. But we made also a few other changes. In particular we study a model with 
parallel updating in discrete time, while a model with continuous time (the 
tricritical `contact process' \cite{Hinrich} instead of TDP proper) was studied 
in \cite{Lubeck}.  These two models should be in the same universality class, 
but a check would be welcome -- and simulating discrete time processes is often 
faster. Finally, while the exponents observed in \cite{Lubeck} were rather close 
to the mean field values, some of them deviated from mean field theory in the
opposite direction of that predicted by the $\epsilon-$expansion. Simulating
a different model in the same universality class with emphasis on different
aspects might clarify whether this indicates a failure (or slow convergence)
of the $\epsilon$-expansion, or hints at problems with the simulations.

In this paper, we study a fully discrete model which is actually a generalization 
of the well known Domany-Kinzel (DK) model \cite{DK}) to 2+1 dimensions.
This generalization to higher dimensions is necessary, because there is 
no non-trivial tricritical point in 1+1 dimensions (the cross over to the 
first order transition, called ``compact DP" \cite{Essam} in this case,
happens at the boundary of the DK phase diagram, just as the phase transition 
in the 1-d Ising model occurs at $T=0$ \cite{footnote1}.
At the tricritical point we 
obtain rather precise estimates of the (tri-)critical exponents, except for 
the cross-over exponent which is affected by large corrections to scaling.
Beyond this point, when the active/adsorbed transition is first order, 
we find that clusters starting with single sites survive with a stretched 
exponential probability. This is similar to the decay of clusters in the 
Grassberger-Chat\'e-Rousseau (GCR) model with re-infection easier than first 
infection \cite{GCR,Del-Hin}, and has a similar qualitative reason \cite{footnote2}.
The boundary between active and adsorbed regions behaves, near the first 
order transition, like a generic fluctuating non-equilibrium (KPZ \cite{KPZ}) 
surface and shows the same scaling laws.

In the next section we will define the model and describe some special 
features of the simulations. Predictions to compare our simulations with 
are reviewed in Sec.3. Results will be presented in Sec.4, while 
the paper end with a discussion in Sec.5.

\section{The generalized Domany-Kinzel model}

The standard DK model lives on a square lattice. Usually, this lattice is 
drawn as tilted by 45$^o$, and a site can be wetted by its two lower neighbours.
For a more easy later generalization to 2+1 dimensions, we use a non-tilted 
lattice, so that site $(x,t+1)$ can be wetted by sites $(x-1,t)$ and $(x+1,t)$. 
The probability to be wetted is $P_1 = y$, if one of these sites was wetted, 
and $P_2=z$, if both were wetted. Here $y$ and $z$ are real numbers between 
0 and 1. If $y=z$, this corresponds to site DP, while bond DP corresponds
to $z=y(2-y)$. If $y<1/2$, any finite cluster of active (=wetted) sites
dies with probability one. For any $y>1/2$, there is a second order phase 
transition (in the DP universality class) to a surviving active phase at 
$z=z_c(y)$. Finally, when $y=1/2$ the cluster dies still with probability 
one, but the average life time is infinite when $z=1$. This is then the case 
of compact DP \cite{Essam}.

In 2+1 dimensions we take a square lattice in space, and each site 
$({\bf x},t+1)$ can be infected by any of the four sites $({\bf x\pm e}_1,t),\; 
({\bf x\pm e}_2,t)$. Assuming still that at least one of these sites has to be
active in order to activate $({\bf x},t+1)$, we have now four wetting 
(``infection", ``activation") probabilities $P_k,\; k = 1,2,3,4$. Site 
percolation corresponds to $P_k = p$ for all $k$, while bond percolation 
corresponds to $P_1 = p$, $P_{k+1} = P_k + (1-P_k)p$ for $k>1$. These 
formulas are easily understood: In site percolation it does not matter how 
many of the neighbours are active, since the site will be wetted anyhow, if 
it can be wetted at all. In bond percolation, the chance to be wetted by 
$k+1$ neighbours is equal to the chance that the first $k$ of them succeeded, 
plus the chance that the last one succeeds if the first were unsuccessful. 
We have thus 4 control parameters, but generically it will be sufficient 
to study a 2-dimensional subspace. We choose (somewhat arbitrarily) the 
following 2-parametric ansatz
\be
   P_1 = p, \quad P_{k+1} = P_k + (1-P_k)pq \;\;\;{\rm for}\;\; k>1       \label{DK3}
\ee
with $0<p<1$ and $q$ such that all $P_k$ are positive and less than 1. This ansatz 
includes both site DP ($q=0$) and bond DP ($q=1$). When $0<q<1/p$, it has a similar 
interpretation to the one given for bond DP: if the first $k$ attempts were all 
unsuccessful in wetting the site, the chance for the next one is not $p$ but is $qp$. 
But Eq.(\ref{DK3}) is independent of this interpretation and is legitimate also
for $q<0$, provided $0\le P_k\le 1$ for all $k$.

Simulations of this model are straightforward. At each time $t$ we have two lists 
$L_{\rm old}$ and $L_{\rm new}$ of sites, together with an integer array S of size 
$L\times L$, where $L$ is the linear size of the system. At the beginning we put 
$t=0$, S and $L_{\rm new}$ are empty, and $L_{\rm old}$ contains the coordinates of 
the seed (we use helical boundary conditions, i.e. each site is labelled by a single 
integer $i$ with $i\equiv i+L^2$, and its four neighbours are $i\pm 1$ and $i\pm L$. 
When going from $t$ to $t+1$, we first go through the list $L_{\rm old}$ and 
increase S[$i$] by one unit, whenever $i$ is a neighbour of a site in $L_{\rm old}$.
At the same time we write each of these $i$ into the list $L_{\rm new}$. In a second
pass, we first erase $L_{\rm old}$, then we check which sites in $L_{\rm new}$ are 
actually wetted (for this, we use the information stored in S). Those who are wetted 
are written into $L_{\rm old}$, and S[$i$] is reset to zero for all checked sites. 
We then erase $L_{\rm new}$, and are ready to go from $t+1$ to $t+2$.

There are two non-trivial tricks for improving this algorithm. The first is needed
in the first order transition regime. There, the survival probabilities decay 
extremely fast with time $t$. Therefore we use enrichment, implemented recursively 
as in the PERM algorithm \cite{PERM}. Essentially, this makes a copy of every 
cluster which survives up to a $t$ where the fraction of surviving clusters is 
below some threshold, and lets the copy and the original evolve independently. 

The second trick is related to histogram reweighting \cite{Dickman,Balle}, but 
it is done on the fly as in 
\cite{Grass03}. Assume a cluster contains $n_k$ sites wetted by $k$ neighbours
($k=1,\ldots 4$) and $b_k$ boundary sites which were {\it not} wetted, although
they had $k$ wet neighbours. Such a cluster is included in the sample with 
probability 
\be
   P(\{n_k\},\{b_k\};p,q) = \prod_{k=1}^4 P_k^{n_k} (1-P_k)^{b_k}.
\ee
For any observable $A$, the average is 
\bea
   \langle A\rangle_{p,q} & = & \sum_{\{n_k\},\{b_k\}}A(\{n_k\},\{b_k\})
                                 P(\{n_k\},\{b_k\};p,q)  \nonumber \\
              & = & M^{-1} \sum_{\rm events} A(\{n_k\},\{b_k\}),
\eea
where the first sum runs over all possible configurations and the second sum 
runs over all $M$ actually simulated clusters. The average for some other pair 
$(p',q')$ of control parameters is then given by 
\be
   \langle A\rangle_{p',q'} = M^{-1} \sum_{\rm events} A(\{n_k\},\{b_k\})
                {P(\{n_k\},\{b_k\};p',q')\over P(\{n_k\},\{b_k\};p,q)}.
      \label{reweight}
\ee
The ratio of probabilities is actually calculated by multiplying the 
corresponding ratios of $P_k$ resp. $1-P_k$ for each site which is wetted
(resp. not wetted) during the build-up of the cluster.
 
This is useful, if we want to estimate several averages during the same run.
It is particularly useful, if we want to estimate derivatives of 
$\langle A\rangle_{p,q}$ with respect to $p$ or $q$. The latter is needed for 
checking scaling laws, as we will discuss below. Such derivatives can be 
obtained directly from Eq.(\ref{reweight}), e.g.
\be
   {\partial \over \partial p} \langle A\rangle_{p,q} = M^{-1} \sum_{\rm events} A(\{n_k\},\{b_k\}) W
\ee
with
\bea
   W & = & \sum_{{\rm wetted} \;\;{\rm sites}} {\partial \over \partial p}\ln P_k(p,q) \nonumber \\
     & + & \sum_{{\rm non-wetted}\;\;{\rm sites}}
           {\partial \over \partial p}\ln (1-P_k(p,q)).
\eea

We made several types of runs. Most extensively, we started from a single seed 
and measured the averages of the number $N(t)$ of sites wetted at time $t$, of the 
survival probability $P(t)$, and of the rms. distance $R^2(t)$ of wetted sites from
the seed \cite{Torre}. In several of these runs we also calculated reweighted 
averages of $N(t)$ and/or $\partial \langle N(t)\rangle/\partial p$. In addition, we made also 
some simulations with fully active initial conditions, where we measured the decay 
of the density $\rho(t)$ of sites active at time $t$. Finally, we made some runs also
with initial conditions where half of the lattice was fully active and the other half
was empty \cite{Lubeck}. In that case one has two phase boundaries, and one can watch 
how these boundaries evolve with time.

In all cases, lattice sizes were sufficiently large that we have no finite size 
corrections. This is most easily checked in single-seed runs: One simply has to check
that the wetted region never hits the boundary of the lattice. In the other cases 
the check is less straightforward, but we always verified that $L/2 \gg t^{1/z}$, 
where $z$ is the dynamical exponent. In most cases this implied $L>1000$ for 
$t\approx 40,000$.

\section{Theoretical predictions}

\begin{figure}
   \begin{center}
     \psfig{file=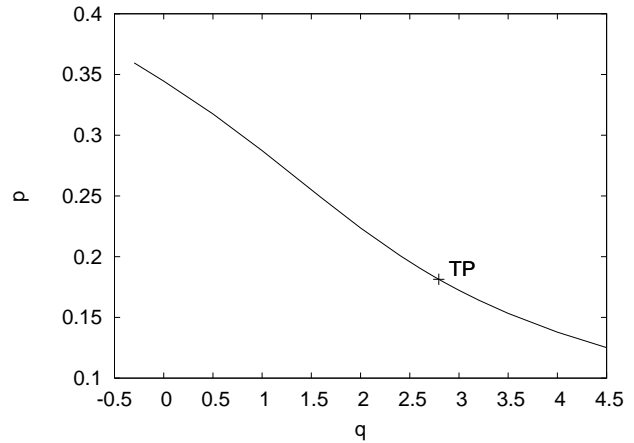,width=6.cm,angle=270}
   \caption{Phase diagram for the generalized DK model in 2+1 dimensions. Below the 
     transition line, activity dies out, while an active phase exists above it. 
     The transition is in the DP universality class to the left of the tricritical
     point marked ``TP", while it is first order to its right. The actual locations 
     of the curve and of the tricritical point were obtained from numerical 
     simulations.}
\label{Fig.phases}
\end{center}
\end{figure}

For each value of $q$, we expect that there is a critical value $p_c(q)$ such that
the activity dies out for $p<p_c(q)$, while $\lim_{t\to\infty}P(t)>0$ 
for $p>p_c(q)$. For all values of $q$, $p_c(q)$ is a decreasing function. Known 
numerical values are for $q=0$ and $q=1$: $p_c(0) = 0.34457(1)$ (site DP 
\cite{Grass-Zhang}) and $p_c(1) = 0.2873381(1)$ (bond DP \cite{Grass-Zhang,Perlsman}).
The phase diagram should thus look roughly as in Fig.~\ref{Fig.phases}.
At the tricritical point TP: $(p,q)=(p^*,q^*)$, the observables $N(t), P(t), 
R^2(t)$, and $\rho(t)$ should follow power laws
\bea
   N(t) \sim t^\eta(1 + O(t^{-\Delta})), \;\;\;
   P(t) \sim t^{-\delta'}(1 + O(t^{-\Delta})), \nonumber \\ 
   R^2(t) \sim t^{2/z}(1 + O(t^{-\Delta})), \;\;\; \rho(t) \sim t^{-\delta}(1 + O(t^{-\Delta})).
                   \label{nprr-scaling}
\eea
Here, $\Delta$ is the exponent of the leading correction to scaling term. 

The upper critical (spatial) dimension is $d_c=3$. For $d>d_c$ one has the mean 
field exponents $\eta=0, \delta' = 1, z = 2$, and $\delta = 1/2$.
The predictions of the $\epsilon-$expansion with $\epsilon = 3-d$ are \cite{Ohtsuki,Janssen04}
\bea
   \eta = -0.0193\epsilon +O(\epsilon^2) \approx -0.019                     \nonumber \\
   \delta' = 1 - 0.0193\epsilon +O(\epsilon^2) \approx 0.981          \nonumber \\
   z = 2+0.0086\epsilon +O(\epsilon^2) \approx 2.009                \nonumber \\
   \delta = {1\over 2} - 0.468\epsilon +O(\epsilon^2) \approx 0.032     
                   \label{epsi}
\eea
Here the numerical values are those obtained for $d=2$ by simply neglecting 
all terms higher than linear in  $\epsilon$. These values satisfy, up to terms 
$O(\epsilon^2)$, the hyperscaling relation \cite{Mendes,Hinrich}
\be
   \eta+\delta+\delta' = d/z.     \label{hyper}
\ee
To derive the latter, remark that $P(t)$ is the probability that there exists 
at least one path of active sites connecting the site $({\bf x}_0,t_0) = (0,0)$ 
to any of the sites $({\bf x},t)$. Similarly, $\rho(t)$ is the chance that there
exists at least one path connecting any site $({\bf x},0)$ to $(0,t)$. The 
probability that $(0,0)$ is connected to $(0,t)$ scales, according to our Ans\"atze
for $N(t)$ and $R^2(t)$, as $P_{\rm conn}\sim N(t)/R^d \sim t^{\eta-d/z}$. On 
the other hand, if there is basically at most one percolating cluster near 
${\bf x}=0$ (an assumption which breaks down for $d>d_c$), and if the site 
$({\bf x},0)$ belongs to it, then the condition for $(0,t)$ to be wetted is 
the same as that for $(0,t)$ to be connected to $(0,0)$. Thus $P_{\rm conn}\approx 
P(t)\rho(t)$, which gives immediately Eq.~(\ref{hyper}).

In the vicinity of the tricritical point one should distinguish between the 
behaviour on the critical curve (but with $q\neq q^*$) from the behaviour 
off the critical curve. For the former, we expect for each observable a 
scaling ansatz with scaling variable $(q^*-q)t^x$, e.g. for $N(t)$ we have
\be
   N(t; q,p = p_c(q)) = t^\eta F((q^*-q)t^x).        \label{Nq}
\ee
Notice that it would be more standard to use $(q^*-q)^{1/x} t$ as scaling 
variable, but then the scaling function replacing $F(z)$ would be singular 
at the origin. The advantage of Eq.~(\ref{Nq}) is that $F(z)$ is analytic
at $z=0$. On the other hand, for $q=q^*$ and $p\neq p_c(q^*)$ we have
\cite{footnote3}
\be
   N(t; q^*, p) = t^\eta G((p_c(q^*)-p) t^y).          \label{Np}
\ee
Again, the standard way of writing this would be in terms of the scaling
variable $(p_c(q^*)-p)^{\nu_\|} t$ with $\nu_\| = 1/y$, the advantage of 
Eq.~(\ref{Np}) being that $G(z)$ is analytic at $z=0$. 

Eq.~(\ref{Np}) shows that the correlation time scales as $\tau \propto
(p_c(q^*)-p)^{\nu_\|}$, and therefore the correlation length scales as 
$\xi \propto (p_c(q^*)-p)^{\nu_\perp}$ with $\nu_\perp = \nu_\|/z$. The cross
over exponent $\phi$ is finally defined as 
\be
   \phi=x/y.
\ee
The $\epsilon-$expansion gives \cite{Ohtsuki,Janssen04}
\bea
   \nu_\| & = & 1+0.0193 \epsilon + O(\epsilon^2) \approx 1.019, \nonumber \\
   \phi & = & {1\over 2} -0.0121\epsilon \approx 0.488,   \nonumber \\ 
   \nu_\perp & = & {1\over 2} + 0.0075\epsilon \approx 0.507.
\eea

Since the tricritical point TP is a fixed point of the renormalization group 
flow which is instable in both directions, we should slowly cross over to the 
scaling laws for normal DP, if we are on the transition line with $q<q^*$. 
Thus, normal DP asymptotics is expected on the entire transition line left of TP.
For DP one has the same scaling laws Eq.(\ref{nprr-scaling}), but with different
exponents \cite{Grass-Zhang,Voigt-Ziff,Perlsman}: $\eta = 0.2303(4), 
\delta=\delta'=0.4509(5),$ and $z = 1.7666(10)$. These values satisfy the 
same hyperscaling relation Eq.~(\ref{hyper}). 

The identity $\delta=\delta'$
follows from a special time reversal symmetry which holds for bond and site
DP, but not for the DK model in general. As we said above, both $P(t)$ and 
$\rho(t)$ are probabilities that one point at the boundary of a time slice
$0\le t'\le t$ is connected to some point on the opposite boundary. For 
bond DP one sees easily that both are the same, i.e. $\rho(t) = P(t)$ exactly.
For site DP one gets the same, if one assumes also all sites with $t'=0$ and 
$t'=t$ as wettable with probability $p$, which gives e.g. $P(1)=\rho(1) = 
p[1-(1-p)^4]$. No relation like that holds for the general DK model.

On the part of the transition line with $q>q^*$ one has compact clusters 
with only few small holes and a slowly moving rough surface. This is analogous
to critical compact DP, except that the latter occurs (in 1+1 dimensions)
only at a single point and has clusters without any holes at all. Mean field
arguments \cite{Lubeck} suggest that the transition is first order along this 
part, in the sense that the density of active sites in the stationary state 
is discontinuous. The growth of a cluster is then similar to the growth of 
a droplet in a usual first order phase transition, except that now one of 
the two phases is not fluctuating. 

Consider now an initially straight 
interface between the two phases. If $p<p_c(q)$, this interface will move 
into the direction of the active phase, i.e. the inactive phase will win.
The opposite is true for $p>p_c(q)$, i.e. the value of $p_c(q)$ is fixed by 
the condition that neither phase will win in the long time limit 
\cite{Schlogl,Ziff-Bros,Lubeck}. Notice, however, that this does not mean
that the interface at $p=p_c(q)$ will not move during short times. Due to 
the asymmetry between the two phases, we expect the generic behaviour of 
non-equilibrium (``growing") interfaces to apply, i.e.  the velocity of the 
critical interface should scale with time as $v \sim t^{-2/3}$ for large $t$
\cite{KPZ}.

Let us finally discuss the growth of a cluster starting from a single site
for $q>q^*, p = p_c(q)$. As in the Grassberger-Chat\'e-Rousseau model with 
negative partial immunization, it is easier to survive for the cluster in its 
already occupied region than to spread out from this region. Although the 
details of the two models are different, we might thus expect that we find also 
in the present model a stretched exponential law, $P(t)\sim \exp(-ct^\alpha)$, 
for basically the same qualitative reasons as in the  Grassberger-Chat\'e-Rousseau 
model \cite{Del-Hin}. We have however no detailed prediction on the exponent $\alpha$.

\section{Results}

\subsection{Gross features of the phase diagram}

\begin{figure}
   \begin{center}
     \psfig{file=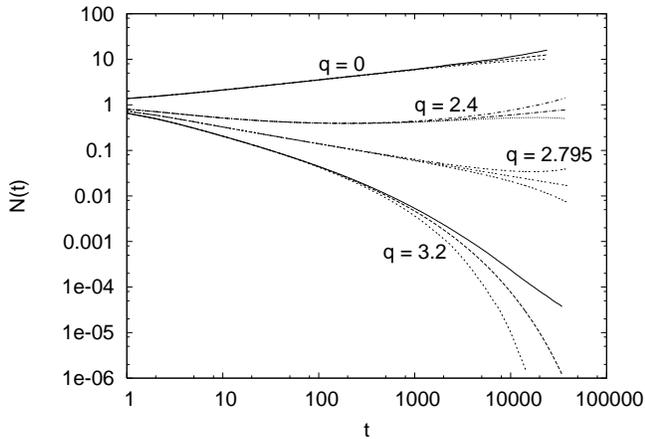,width=6.cm,angle=270}
   \caption{Log-log plots of $N(t)$ for four different values of $q$, and for three
     values of $p$ for each $q$. The three $p-$values are chosen to be subcritical,
     critical, and supercritical (bottom to top). Within each triple, they differ by 
     less than $0.05\%$. The four $q-$values are DP, 
     crossover from TDP to DP, TDP, and first order (top to bottom).}
\label{Nt-rough}
\end{center}
\end{figure}

\begin{figure}
   \begin{center}
     \psfig{file=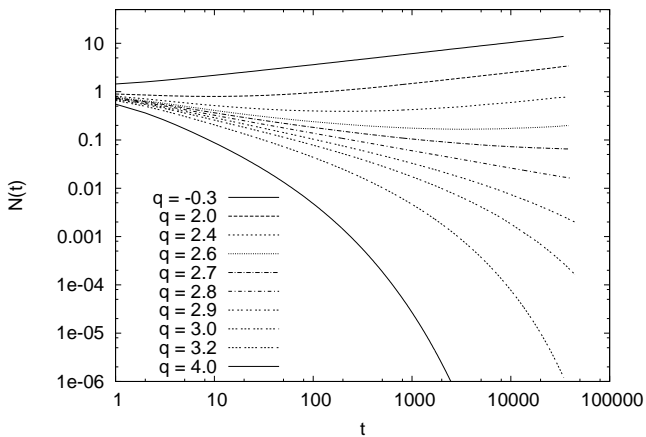,width=6.cm,angle=270}
   \caption{Log-log plots of $N(t)$ for several values of $q$, each at the best 
     estimate of $p_c(q)$.}
\label{crit}
\end{center}
\end{figure}

For ordinary DP, the observable most sensitive to the precise value of $p_c$ is 
$N(t)$. We expect therefore that $N(t)$ is also the best observable to locate the 
tricritical point. In Fig.~\ref{Nt-rough} we show $N(t)$ for twelve pairs of 
control parameters $(p,q)$. These represent three values of $p$ (critical, sub-, and 
supercritical) for each of four values of $q$. The latter are chosen to be: (i) site
DP ($q=0$); (ii) in the crossover region from TDP to DP ($q=2.4$); (iii) close to TDP 
($q \approx q^* \approx 2.8$) and (iv) $q>q^*$. First of all, we see from Fig.~\ref{Nt-rough} 
that there are indeed two straight lines. These are of course the candidates for 
ordinary and for tricritical DP. Secondly, we see that we can determine $p_c(q)$, for 
each value of $q$, with rather high precision. For $q=q^*$ we just have to look 
for a power law, just as for $q\ll q^*$. For $q$ in the cross-over region, i.e. 
close to $q^*$ but not at $q^*$, the determination of  $p_c(q)$ is less easy.
But we still can get rather precise estimates for $q$ slightly below $q^*$, if
we assume that the slopes of the critical curves approach the slope $\eta = 0.2303$ 
of the critical DP curve monotonously. For $q\gg q^*$, finally, it is 
more easy to determine $p_c(q)$ from the requirement that the velocity of a 
straight interface scales as $t^{-2/3}$ (see details below) than from the 
behaviour of $N(t)$. Numerical results for $p_c(q)$ obtained in this way are 
collected in Table 1. Several critical curves are plotted again in Fig.~\ref{crit}.

 \begin{table}
 \begin{center}
 \caption{Estimates of $p_c(q)$.}
   \label{table1}
 \begin{ruledtabular}
 \begin{tabular}{lcl}
 $q$ &        $p_c(q)$      &  comment \\
\hline

 -0.3  & 0.359568(3) &                 \\
  0.0  & 0.344575(5) &     site DP     \\
  0.5  & 0.317505(3) &                 \\
  1.0  & 0.287339(2) &     bond DP     \\
  1.6  & 0.248648(2) &                 \\
  2.0  & 0.223647(2) &                 \\
  2.40 & 0.200939(1) &                 \\
  2.60 & 0.1906655(9) &                \\
  2.70 & 0.185809(1) &                 \\
  2.75 & 0.1834507(9) &                \\
  2.78 & 0.1820572(7) &                \\
  2.79 & 0.1815965(4) &                \\
  2.795& 0.1813672(3)  &   TDP         \\
  2.80 & 0.1811382(4) &                \\
  2.82 & 0.180225(1) &                 \\
  2.85 & 0.178870(2) &                 \\
  2.9  & 0.176648(5) &                 \\
  3.0  & 0.172337(5) &                 \\
  3.2  & 0.164243(4) &                 \\
  3.5  & 0.153313(3) &                 \\
  4.0  & 0.137830(3) &                 \\
  4.5  & 0.12508(1) &                 \\
  5.0  & 0.11444(1) &                 \\
 \end{tabular}
 \end{ruledtabular}
 \end{center}
\end{table}

\begin{figure}
   \begin{center}
     \psfig{file=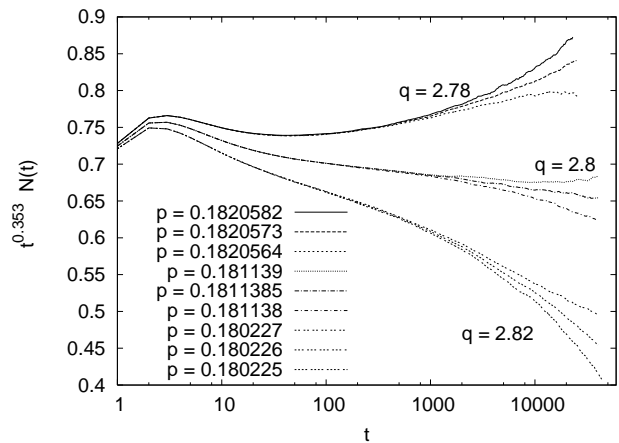,width=6.cm,angle=270}
   \caption{Log-linear plots of $t^{-\eta} N(t)$ for several values of $q\approx q^*$
     and $p\approx p_c(q)$, with $\eta = -0.353$.}
\label{N_tricrit}
\end{center}
\end{figure}

\subsection{Scaling at the tricritical point}

Plots like those in Fig.~\ref{Nt-rough} are of course not suitable for precise 
determinations of critical parameters. For a closer look at the tricritical region, 
we show in Fig.~\ref{N_tricrit} the product $t^{-\eta} N(t)$ with $\eta = -0.353$. 
This value of $\eta$ is our best estimate, i.e. the tricritical curve should be 
horizontal if there were no corrections to the scaling behaviour. Actually we see 
that there are quite strong corrections, manifesting themselves as a bump at $t\approx 3$.
The data shown in Fig.~\ref{Nt-rough} and similar data at other 
near-by values of $q$ show that 
\be
   q^* = 2.792(6),\;\;\; p_c = 0.1831534(5) + 0.458(2.792 - q^*),
\ee
and
\be
   \eta = -0.353(9).
\ee
Although the last estimate has the same sign as the prediction from the $\epsilon-$expansion,
it is larger by nearly a factor 20!

Once we have fixed the tricritical point, we can immediately obtain $\delta'$ and $z$ 
from the scaling of $P(t)$ and $R^2(t)$. We do not show the corresponding data. The 
estimates found by standard extrapolation methods are
\be
   \delta' = 1.218(7),\;\;\; z = 2.110(6).
\ee
For $z$ we find that the deviation from the mean field value $z=2$ has the same sign
as predicted (TDP spreads subdiffusively, in agreement with the intuitive notion that
spreading should be slowed down compared to ordinary DP). But again the difference from
mean field is an order of magnitude larger than predicted. For the correction to $\delta'$, 
the $\epsilon-$expansion predicted even a wrong sign.

Finally, in order to measure the exponent $\delta$, we made runs with fully active 
initial conditions. Again we do not show the raw data, as they contained little 
surprises. They give
\be
   \delta = 0.087(3).
\ee
In spite of the huge difference from mean field behaviour (where $\delta=1/2$), this 
is in surprising good agreement with Eq.~(\ref{epsi}). The hyperscaling relation is 
very well satisfied with these estimates,
\be
   \eta + \delta + \delta'  - d/z = 0.004(13),
\ee
showing that they are at least internally consistent.

\subsection{Scaling near the tricritical point}

\begin{figure}
   \begin{center}
     \psfig{file=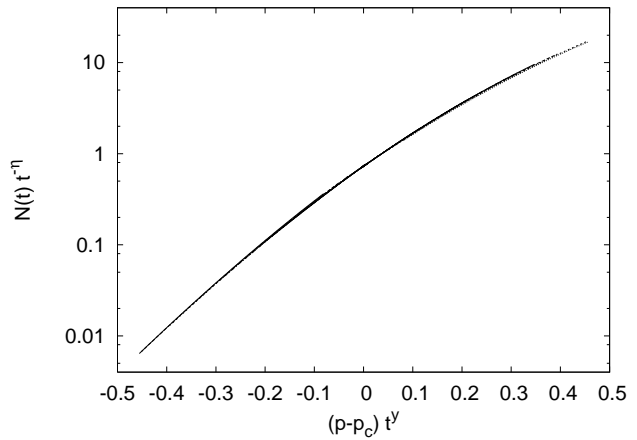,width=6.cm,angle=270}
   \caption{Log-linear plots of $t^{-\eta} N(t)$ against $(p_c(q^*)-p) t^y$, with 
     $q^*=2.8, \eta = 0.362,$ and $y = 0.87$. Only points with $t\geq 30$ are shown. 
     Values of $p$ range from 0.17693 to 0.185350. Notice the seemingly perfect 
     data collapse in spite of the slightly wrong tricritical parameters and the 
     scaling violations seen in Fig.~\ref{N_tricrit}.}
\label{nu-collapse}
\end{center}
\end{figure}

A popular method to find correlation length and cross-over exponents are data collapse
plots. In view of a scaling law like Eq.~(\ref{Np}), it seems natural to plot $N(t)/t^\eta$
against $(p_c(q^*)-p) t^y$ and determine $y$ (or, equivalently, $\nu_\|$) so that all data fall onto a single 
curve. Such a plot is shown in Fig.~\ref{nu-collapse}, where we have chosen $y = 0.87$.
It seems to give a perfect collapse, i.e. there are no indications of any scaling 
violations. This might seem strange in view of the rather strong violations seen in 
Fig.~\ref{N_tricrit}. It is true that we suppressed them by plotting only data with 
$t>30$, but this should not eliminate them completely. Rather, they are not seen in 
Fig.~\ref{nu-collapse} because of the coarse scale on the y-axis, and because we also 
changed slightly the tricritical parameters. We 
used $q^*=2.8$ and $\eta = 0.362$, while the plot would be definitely less clean if our
best estimates had been used.

\begin{figure}
   \begin{center}
     \psfig{file=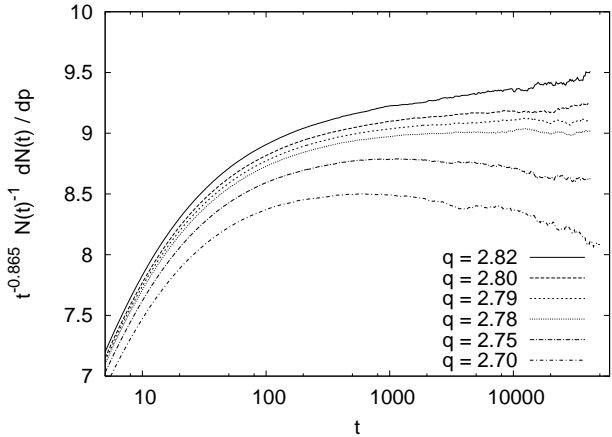,width=6.cm,angle=270}
   \caption{Log-linear plots of $t^{-y} \partial \log N(t)/\partial p$ against $t$.}
\label{nu}
\end{center}
\end{figure}

A much more reliable and precise method consists in measuring $\partial N(t;q,p)/\partial p$. 
As we have pointed out in Sec.~2, this can be done straightforwardly. From the scaling 
ansatz Eq.~(\ref{Np}) we have 
\be
   {\partial \log N(t;q^*, p_c(q^*)) \over \partial p} = t^y {d\log G(z)\over dz}|_{z=0} \propto t^y,
\ee
up to corrections to scaling. This quantity was found to depend very weakly on the precise
value of $p$. Thus the main uncertainty comes from its dependence on $q$, together with
our rather large error for $q^*$. We plot therefore in Fig.~\ref{nu} the quantity $t^{-y} \partial \log N(t)/\partial p$ 
for various values of $q$ close to $q^*$, taking for each curve our best estimate of $p_c$.
We see quite strong corrections to scaling (as we expected), but they do not prevent 
us from a very precise estimation of $y$. Fitting with a correction to scaling exponent
$\Delta =0.66$, i.e. with an ansatz $\partial \log N(t)/\partial p \propto t^y(1+{\rm const}/t^{0.66})$, 
we obtain 
\be
   y = 0.865(3),\;\;\; \nu_\| = 1.156(4),
\ee
and, using the previously determined value of $z$, $\nu_\perp = 0.547(3)$. Using this, 
we obtain also $\beta=\nu_\| \delta = 0.100(4)$ and $\beta'=\nu_\| \delta' = 1.408(10)$,
where $\beta$ and $\beta'$ are defined via the scalings $\lim_{t\to\infty}\rho(t)\sim 
(p-p_c(q^*))^\beta,  \lim_{t\to\infty} P(t) \sim (p-p_c(q^*))^{\beta'}$, both for $p>p_c$.
Again these estimates, albeit being close to their mean field values, deviate from 
them much more than predicted by the $\epsilon-$expansions (which is particularly simple 
for $\beta'$, which should be equal to $1+O(\epsilon^2)$).

Finally, in order to measure the cross-over exponent $\phi$, we can either 
use again a data collapse plot, or we can try to estimate the derivative of 
$N(t;q,p)$ (or of any other observable) along the curve $p=p_c(q)$. The latter
seems however not easy. It is of course straightforward to measure the 
derivative in any given direction. But the orientation of the transition 
line $p=p_c(q)$ is only approximately known, and the derivative should depend
strongly on this orientation. Thus we determined $\phi$ only from its data
collapse, in spite of the shortcomings mentioned above.

\begin{figure}
   \begin{center}
     \psfig{file=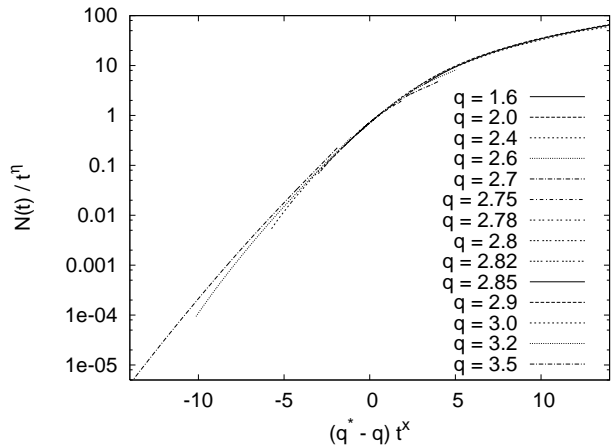,width=6.cm,angle=270}
   \caption{Log-linear plots of $t^{-\eta} N(t;q,p_c(q))$ against $(q^*-q) t^x$, with
     $q^*=2.792, \eta = 0.353,$ and $x = 0.31$. Only points with $t\geq 24$ are shown.
     Values of $q$ range from 1.6 to 3.5.}
\label{phi-collapse}
\end{center}
\end{figure}

Indeed, the situation is now much worse than in the data collapse plot for 
$\nu_\|$, partly because of the uncertainties in $p_c(q)$. In Fig.~\ref{phi-collapse} 
we show the values of $t^{-\eta} N(t;q,p_c(q))$ plotted against $(q-q^*)t^x$. 
The values of $\eta$ and of $p_c(q)$ are as determined above, and $x = 0.31$. 
The data collapse is now much worse than in Fig.~\ref{nu-collapse}.This may 
be due to the fact that Eq.~(\ref{Nq}) has even larger corrections to scaling 
than Eq.~(\ref{Np}), but it may also be due to errors in estimating $p_c(q)$. The
data collapse would improve considerably, if we would shift $p_c(q)$ for $q<q^*$
systematically towards higher values. In that case the critical curves $\ln N(t)$
versus $\ln t$ would no longer be concave, i.e. $d\ln N(t)/d\ln t$ would no 
longer approach approach the value $\eta_{\rm DP} = 0.2303$ monotonically from 
below. Although we cannot rule out such a behaviour, we prefer to keep the 
estimates of $p_c(q)$ and to increase the estimated error of $\phi$. We thus 
obtain $x = 0.31(3)$ and 
\be
   \phi = x\nu_\| = 0.36(4).
\ee
Again the deviation from the mean field value ($\phi = 1/2$) is in the same 
direction as predicted by the $\epsilon-$expansion, but is much larger in 
absolute value.

The numerical values of the tricritical exponents, together with the previous estimates 
from the $\epsilon-$expansion and from the simulations in Ref.~\cite{Lubeck},
are collected in Table 2.

 \begin{table*}
 \begin{center}
 \caption{Estimated critical exponents for tricritical DP in 2+1 dimensions. Values for the 
    spatial fractal dimension $D_f$ and for the exponents $\gamma,\sigma,$ and $\eta_\perp$ 
    not discussed in the text were obtained by means of scaling and hyperscaling relations
    as indicated in column $\#1$. They are included here for easier comparison with the 
    simulations of Ref.~\cite{Lubeck} (last column).}
   \label{table2}
 \begin{ruledtabular}
 \begin{tabular}{lcrcc}
   & defined in &present work& $\epsilon-$expansion \cite{Ohtsuki,Janssen04} & Ref.~\cite{Lubeck} \\
\hline
 $\eta$                 & Eq. (11)   & $-0.353(9)$    & $   -0.0193\epsilon = -0.019$ &   --   \\
 $\delta$               & Eq. (11)   & $ 0.087(3)$    & $1/2-0.4677\epsilon =  0.032$ &   --   \\
 $\delta'$              & Eq. (11)   & $ 1.218(7)$    & $1 - 0.0193\epsilon =  0.981$ &   --   \\
 $z$                    & Eq. (11)   & $ 2.110(6)$    & $2 + 0.0086\epsilon =  2.009$ &   --   \\
 $\beta = \nu_\|\delta$ &  --        & $ 0.100(4)$    & $1/2-0.4580\epsilon =  0.042$ & $0.14(2)$  \\
 $\beta'= \nu_\|\delta'$&  --        & $ 1.408(10)$   & $1 + {\cal O}(\epsilon)$      &   --   \\
 $\nu_\|$               & Eq. (15) ff& $ 1.156(4)$    & $1 + 0.0193\epsilon =  1.019$ &   --   \\
 $\nu_\perp = \nu_\|/z$ &  --        & $ 0.547(3)$    & $1/2+0.0075\epsilon =  0.507$ & $0.59(8)$  \\
 $D_f = d-\beta/\nu_\perp$&  --      & $ 1.817(8)$    & $2 - 0.0690\epsilon =  1.931$ & $1.76(5)$  \\
 $\gamma=\nu_\|(1+\eta)$&    --      & $ 0.748(11) $  & $1 + {\cal O}(\epsilon)$      & $1.00(6)$  \\
 $\sigma=\gamma+\beta$  &    --      & $ 0.848(12) $  & $1 - 0.0193\epsilon =  0.981$ & $1.12(5)$  \\
 $\eta_\perp = 2-d+2\beta/\nu_\perp$&--& $ 0.366(16)$ & $1 - 0.8620\epsilon =  0.138$ & $0.42(24)$ \\
 $\phi$                 & Eq. (16)   & $ 0.36(4)$     & $1/2-0.0121\epsilon =  0.488$ & $0.55(3)$  \\
 \end{tabular}
 \end{ruledtabular}
 \end{center}
\end{table*}

\subsection{The first order transition}

As we said, data like those shown in Figs.~\ref{Nt-rough},\ref{crit} are strongly suggestive of
stretched exponentials, but it is notoriously difficult to obtain the precise asymptotic 
behaviour from such curves. It seems not even clear which of the curves for $q=3.2$ in 
Fig.~\ref{Nt-rough} is closest to the critical one. An alternative way to obtain the 
critical point in this case consists in the following. We start with initial conditions
where one half of the square lattice of size $L\times L$ (let us say all sites with 
$L/4 \le x < 3L/4$) is active and the other half is dead. Thus we have two interfaces, at 
$x=L/4$ and at $x=3L/4$ (we assume that $L$ is a multiple of 4, and boundary conditions 
are periodic in the $x$ direction). We then measure the density $\rho(x,t)$ as the system 
evolves. The initial step functions will become smooth sigmoidals, corresponding to fuzzy
interfaces. Their positions are measured by measuring 
\be
   X(t) = {2\sum_{x=0}^{L-1} \rho(x,t) |x-(L-1)/2| \over \sum_{x=0}^{L-1} \rho(x,t)} - L/4.
\ee
Initially, $X(t=0)=0$. For large $t$, $X(t)$ increases linearly with $t$ if $p>p_c(q)$, 
while it decreases if $p<p_c(q)$.

\begin{figure}
   \begin{center}
     \psfig{file=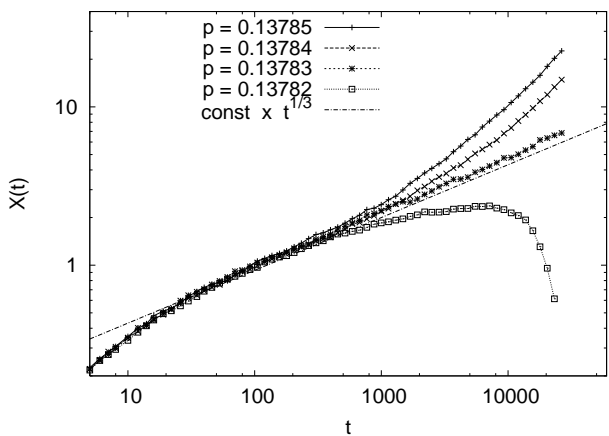,width=6.cm,angle=270}
   \caption{Log-log plots of $X(t)$ versus $t$ for $q=4.0$ and four values of $p$. The 
     straight line corresponds to $X(t)\propto t^{1/3}$, the generic behaviour of a 
     non-equilibrium surface which neither grows nor recedes asymptotically.}
\label{Xt}
\end{center}
\end{figure}

Data for $q=4.0$ are shown in Fig.~\ref{Xt}, where we have plotted $\ln X(t)$ against 
$\ln t$ for four different values of $p$. In addition we show in this figure the 
power law $X(t)\propto t^{1/3}$ which is generically expected for a rough 
non-equilibrium surface which neither grows nor recedes for $t\to\infty$. We see that 
this is indeed the asymptotic behaviour for $p= 0.137830(3)$, which is thus our best 
estimate for $p_c(4.0)$.

Similar behaviour is found also for other values of $q>q^*$, except when $q$ is very 
close to $q^*$. There the transient behaviour seen in Fig.~\ref{Xt} for $t<20$ extends 
to much larger $t$, reflecting the increased internal fuzziness of the interface.

\begin{figure}
   \begin{center}
     \psfig{file=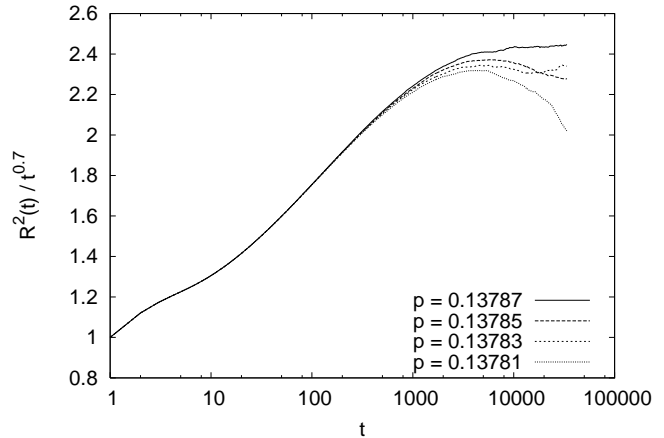,width=6.cm,angle=270}
   \caption{Log-linear plots of $R^2(t) / t^{0.7}$ versus $t$, for $q=4.0$ and 
     four values of $p$.}
\label{R-first}
\end{center}
\end{figure}

\begin{figure}
   \begin{center}
     \psfig{file=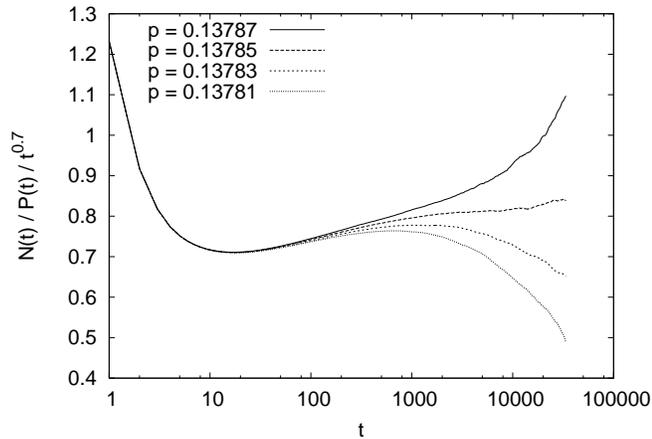,width=6.cm,angle=270}
   \caption{Log-linear plots of $N(t)/P(t) / t^{0.7}$ versus $t$, for $q=4.0$ and
     four values of $p$.}
\label{N-first}
\end{center}
\end{figure}

It is an interesting question how clusters starting from a single site grow at the 
first order transition line. A priori one might expect the behaviour to be similar to 
that of a small droplet in a gas at boiling temperature, or a small domain of inverted 
spins in an Ising magnet at $T<T_c$. But these analogies might break down due to the 
essentially non-equilibrium nature of DP. Let us concentrate again on $q=4.0$ (at other 
values of $q$, the behaviour was similar). Quantities $R^2(t)$ and $N(t)/P(t)$ (i.e., the 
average size of {\it surviving} clusters) are plotted against $t$ in Figs.~\ref{R-first}
and \ref{N-first}. Actually, in order to see the scaling more clearly, in both cases 
the plotted quantity was first divided by a suitable power of $t$, so that the curves
would be straight horizontal lines if there were pure power laws. But the curves, 
in particular those for the critical value $p=0.13783$, are far from straight. They
are compatible with approximate scaling laws $R^2 \sim N(t)/P(t) \sim t^{0.5}$ to $t^{0.6}$,
but we are still far from the asymptotic regime. For $t\to\infty$ we should expect 
that surviving clusters are roughly circular and have a compact interior with density
$\rho = N(t) / P(t) / R^d(t)$. If the transition is indeed first order, this density
should tend to a positive constant for $t\to\infty$. As seen from Figs.~\ref{R-first}
and \ref{N-first} this is indeed the case, although this limit is reached rather late.


\section{Discussion}

The main result of the present analysis is the rather poor agreement with the 
$\epsilon$-expansion. This is somewhat surprising, since the upper critical dimension
is $d_c=3$ for this problem, and therefore terms linear in $\epsilon$ should give
a rather good description in two dimensions. It is even more strange, since the 
contributions of order $\epsilon$ are extremely small for all critical exponents -- 
except for $\delta$, and it is only for $\delta$ that there is fairly good agreement.
At least, for most exponents (except $\delta'$) the deviations from the mean field 
predictions have the same sign as predicted to order $\epsilon$.

Our results are also in rather poor agreement with those of the simulations of 
L\"ubeck \cite{Lubeck}. E.g., he found $\phi=0.55(3)$ which is $5\sigma$ from our
estimate, and which deviates from the mean 
field value $1/2$ in the opposite direction than the $O(\epsilon)$ term. His estimate 
$\nu_\perp = 0.59(8)$ agrees with ours within the error bars, but his estimate
$\beta = 0.14(2)$ exceeds ours by 2 standard deviations. The reason for this 
discrepancy is not clear. It could be due to the fact that the analysis in 
\cite{Lubeck} was mainly based on data collapse plots, which makes it very difficult 
to take into account possible corrections to scaling.
As we pointed out in Sec.4.C, such plots can hide even large systematic deviations 
from the supposed scaling behaviour.

Overall, the simulations presented in this paper produced hardly any difficulties or 
surprises. Except for the 
data collapse for the cross-over exponent $\phi$, all corrections to scaling were 
rather modest. We thus believe that our analysis is robust and does not hide any 
large systematic errors. 

This is in marked contrast to the PCPD \cite{Hinrich-Henkel}, which can be viewed
as an alternative tricritical version of standard DP. It would be nice to have a 
model where the PCPD arises as a singular limit of TDP. The version of the generalized
Domany-Kinzel model which we used in the present paper does not give rise to the 
PCPD in any limit.

After having discussed tricritical directed percolation, we should mention also 
tricritical ordinary (undirected) percolation. A
field theoretic renormalization treatment of this problem has been given recently 
by Janssen and coworkers \cite{JMS}. Discrete lattice model versions can be 
formulated in close analogy to the present work. Let us consider the dynamic version,
i.e. the generalized epidemic process (GEP) \cite{GEP}. This is very similar to directed 
percolation viewed as an epidemic process, except that lattice sites which had once
been infected cannot be re-infected again (this is also known as the SIR model, for 
``susceptible-infected-removed"). In a generalized GEP (GGEP) one can modify the 
probabilities $P_k$ to be infected by $k$ infectious neighbours in the same way as we
did in the present paper (Eq.~\ref{DK3}), but one can also use other Ans\"atze for 
$P_k$ with the same qualitative behaviour. Typically, a tricritical regime is reached 
when $P_k$ increases sufficiently fast with $k$. Models of this type were studied 
some time ago by Cieplak, Robbins, Koiller and others \cite{Cieplak}. A comparison of 
their results with the theoretical work of \cite{JMS} would be very welcome. Also, 
there are some surprising claims in 
\cite{Cieplak}, e.g. it is claimed in some of these papers that there is also a 
transition between pinned fractal (i.e. percolation like) clusters and compact clusters 
with self-affine surfaces 
in 1+1 dimensions. One should expect such a transition to occur in higher dimensions, 
but like the transition to compact DP in the 2-d Domany-Kinzel model, it should occur in 
1+1 dimensions only at the boundary of the control parameter space.

Finally, it is instructive to compare different epidemic models, and we shall finish this 
paper with a short review of how the different schematic models discussed in the 
literature are related. Let us first assume we can have three different types of 
individua in a population: Healthy susceptible ones (S), ill and infective ones (X), 
and ``removed" individua (R), which might either mean immune or dead. Then we have 
the following basic schemes:

\begin{itemize}
\item The only reaction is $S+X \to X+X$, i.e. ill individua remain infective forever.
This gives Eden growth \cite{KPZ}.
\item In addition we have $X \to R$, i.e. ill individua have a finite life time and are 
removed thereafter. This gives the GEP.
\item Instead of $X \to R$ we have $X\to S$, i.e. after individua have recovered they 
become again susceptible. This gives DP.
\item If we add a reaction $R\to S$ to the GEP, it should bring the model into the DP
universality class, except when the rate of this last reaction is very small. In the 
limit when this rate is much smaller than all other rates, we obtain the Bak-Chen-Tang
\cite{BCT} forest fire model.
\end{itemize}

If we have in addition a fourth class (W) of individua weakened by the contact with 
an infective, but neither ill nor infective themselves, then we get three more schemes:

\begin{itemize}
\item If we change the recovery in DP to $X\to W$ and add $W+X\to X+X$, i.e. recovered individua
have a different chance to be re-infected by a subsequent contact with ill ones, we obtain 
the model of Grassberger, Chat\'e, and Rousseau \cite{GCR}. Notice that in this model weakened 
individua stay weak forever, and never regain their strength.

\item If we include in addition the reaction $W\to S$, then weakening is only 
transient and we should expect to be again in DP.

\item If, however, transient weakening happens not {\it after} but {\it instead of} an 
infection, i.e. if we add $S+X\to W+X$, $W+X\to X+X$, and $W\to S$ to DP, we obtain the 
model discussed in the present paper.

\item If we add $S+X\to W+X$ and $W+X\to X+X$ (but not $W\to S$) to the GEP, we 
end up at the GGEP.
\end{itemize}

Obviously this list is not exhaustive. It seems however to contain all interesting cases, 
if we assume that the reservoir of susceptible individua is unlimited (conservation of 
the total population size is irrelevant), and that there is some local saturation 
mechanism which prevents any of the densities to explode.
Whether this is indeed true, or whether there exist more such models with interesting novel 
behaviour, is still an open question.

Acknowledgements:

I am indebted to Sven L\"ubeck for showing me his results prior to publication, 
without which I would not have started this work. I also want to thank Maya Paczuski 
for several discussions, and I want to acknowledge the hospitality of the Perimeter
Institute where this work was started. Finally I want to Hannes Janssen and Sven 
Luebeck for several critical e-mails and for carefully reading the paper.


\begin{thebibliography}{}
\bibitem{footnote2} In both models, it is easier for an epidemic to survive in a region, 
   than to spread out into new regions. While this happens by an explicit long time memory 
   effect in the GCR model (sites which were once infected are more easily re-infected), 
   it happens in the present model because sites inside an infected region have more 
   infected neighbours than sites on the outer boundary. In both cases, a critical 
   cluster grows slower than in ordinary DP, and thus has a larger chance to die
   due to random fluctuations.
\bibitem{footnote3} Eq.(\ref{Np}) holds indeed on any smooth curve in the $(p,q)$-plane
   which cuts the transition curve transversally in the tricritical point, not only 
   for the vertical line $q=q^* = {\rm const}$.
\end{thebibliography}
\end{document}